\documentclass{PoS}

\usepackage{geometry}
\newcommand{\beq}{\begin{eqnarray}}
\newcommand{\eeq}{\end{eqnarray}}

\newcommand{\Lams}{\Lambda_{\overline{\rm MS}}}

\newcommand{\be}{\begin{equation}}
\newcommand{\ee}{\end{equation}}
\newcommand{\lwrsim}{\raise0.3ex\hbox{$<$\kern-0.75em\raise-1.1ex\hbox{$\sim$}}}
\newcommand{\lgrsim}{\raise0.3ex\hbox{$>$\kern-0.75em\raise-1.1ex\hbox{$\sim$}}}

\def\C2#1#2{({\cal C}_2)_{#1}^{#2}}

\def\eq#1{eq.~(\ref{#1})}

\def\VEV#1{\langle #1 \rangle}
\title{$\alpha_S$ from Lattice QCD: progresses and perspectives for a realistic full-QCD determination of the running Strong coupling}

\ShortTitle{$\alpha_S$ from Lattice QCD}

\author{B. Blossier, Ph. Boucaud, M. Gravina, O. P\`ene\\
        Laboratoire de Physique Th\'eorique\\
        {Universit\'e de Paris-Sud, B\^atiment 210, 91405 Orsay Cedex,
        France}\\
        E-mail: \email{Benoit.Blossier,Philippe.Boucaud,Mario.Gravina,Olivier.Pene@th.u-psud.fr}}

\author{F. De soto\\
        Dpto. Sistemas F\'isicos, Qu\'imicos y Naturales, 
        U. Pablo de Olavide, 41013 Sevilla, Spain.\\
        E-mail: \email{fcsotbor@upo.es}}

\author{V. Morenas\\
        Laboratoire de Physique Corpusculaire, 
        Universit\'e Blaise Pascal, CNRS/IN2P3, France\\
        E-mail: \email{morenas@in2p3.fr}}

\author{\speaker{J. Rodriguez-Quintero}  \\
        University of Huelva, Dpto. Fisica Aplicada, Fac. CCEE, 21071 Huelva \\
        E-mail: \email{jose.rodriguez@dfaie.uhu.es}}

\abstract{
\begin{center}
\raisebox{5mm}{\hbox{
\includegraphics[width=3cm]{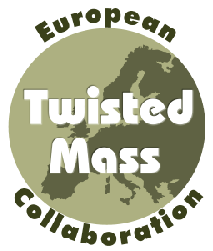}}}
\end{center}

Some very recent computations of $\alpha_{\overline{\rm MS}}(M_Z)$ from $N_f=1+1$ lattice simulations 
and of the running of the Strong coupling, obtained from the lattice ghost-gluon vertex, over a large momentum window 
are very briefly reviewed. 

}

\FullConference{35th International Conference of High Energy Physics\\
		 July 22-28, 2010\\
		 Paris, France}

\begin{document}

\section{Introduction}

$\Lambda_{\overline{MS}}$ is the scale of strong interactions. This parameter has to be 
taken from experiment and can be determined from the running of the QCD coupling constant.
This latter had been calculated in the past by following a variety of non-perturbative ways on the lattice (see ~\cite{Luscher:1993gh,Gockeler:2005rv,Alles:1996ka,Boucaud:tanti,Boucaud:2000ey,Sternbeck:2007br} and references therein) 
from quenched and $N_f=2$ gauge configurations; although those results cannot be properly compared with the experimental 
determinations of the Strong coupling, $\alpha_S(M_Z)$, because of the inaccesibility 
of the threshold scales of $\mu=\mu_{u,d,s}$. 
We will very shortly comment first on the very recent reported progress on the lattice determinantion of $\alpha_S(M_Z)$ 
from $N_f=2+1$ simulations, where perturbation theory is used for the matching at the threshold for the charm mass, 
$\mu=O(1)$ GeV,  from $N_f=3$ to $N_f=4$, implying not to take into account the non-perturbative effects, still important 
at this scale. Then, we will focussed on the study of the running itself of the Strong coupling through the comparison 
between the perturbative and lattice determinations of $\alpha_S$ from the ghost-gluon coupling 
over a large momentum window~\cite{Blossier:2010we}. This has been done from quenched lattice simulations and 
with $N_f=2$ twisted mass quark flavours~\cite{Boucaud:2008xu} and reveals the presence of a 
dimension-two $\VEV{A^2}$ condensate, signaling that momenta considered in lattice simulation are in a 
non-perturbative region. 

\section{$\alpha_{\overline{\rm MS}}(M_Z)$ from the lattice}

There have been very recent estimates of $\alpha_{\overline{\rm MS}}(M_Z)$ by applying different procedures 
(for a recent report, see section 4.1 of~\cite{Aoki:2010yq}) from lattice simulations.
\begin{figure}
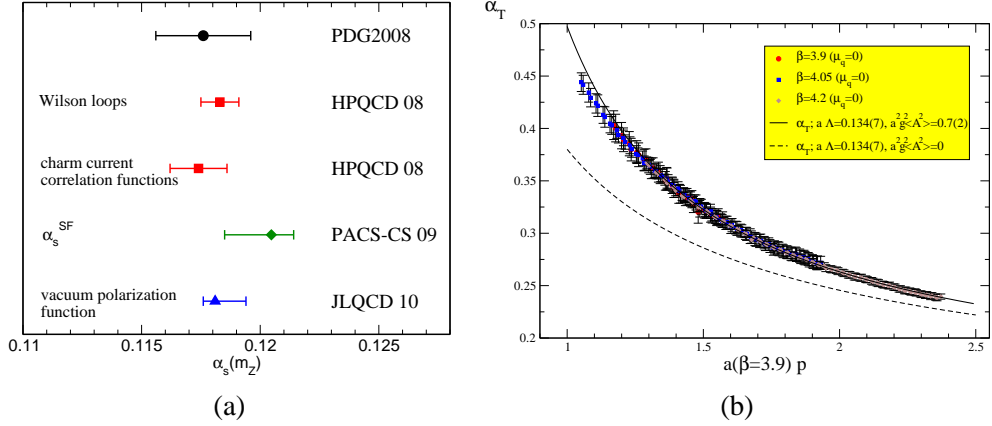

  \begin{center}
  \begin{tabular}{cc}
  \includegraphics[height=5cm]{alpha_s.eps} &
  \includegraphics[height=5cm]{alphaR.eps} \\
  (a) & (b)
  \end{tabular}
  \caption{\small (a) Summary of $\alpha_{\overline{\rm MS}}(M_Z)$ from $n_f=2+1$
   lattice simulations, compared with PDG 2008 average (black)
   \cite{Amsler:2008zzb} discussed in the text. The red points are for determinations using staggered fermions, 
   the green for one using Wilson and the blue overlap fermions.
   The plot is taken from ~\cite{Aoki:2010yq}.
(b) $\alpha_T$ from the lattice, after applying the appropriate lattice-artefacts curing procedure, confronted to the 
continuum formula obtained from PT and including OPE non-perturbative corrections. The solid line is for 
the complete non-perturbative expression, while dotted stands only for the perturbative four-loop one, 
$\alpha_T^{\rm pert}$. The momentum in the x-axis is expressed in lattice units of $a(\beta=3.9)^{-1}$. 
The plot is taken from ref.~\cite{Blossier:2010we}. }
 \label{fig:alpha_s}
  \end{center}
\end{figure}
%---------------

In ref.~\cite{Davies:2008sw}, the coupling $\alpha_V$ defined from Wilson loops 
through lattice perturbation theory is computed and then matched to
$\alpha_{\overline{\rm MS}}(\mu)$ at three-loop. The authors of ref.~\cite{Allison:2008xk} 
use the continuum three-loop expression of a moment of charm
current-current correlation function and get $\alpha_{\overline{\rm MS}}$ by 
comparison with the lattice estimates of the same moment. The work of 
ref.~\cite{Aoki:2009tf} use step scaling of the SF coupling, where the renormalization scale is
set from the inverse linear lattice extension $\mu=1/L$, and matching to $\overline{\rm MS}$ 
with three-loops PT. Finally, in ref.~\cite{Shintani:2010ph}, 
the continuum vacuum polarization function has been obtained through
operator product expansion and the relevant coefficients has been
calculated up to three-loops, the renormalization scale being set from the size
of the injected momentum at the current. The results of $\alpha_{\overline{\rm MS}}(M_Z)$ estimates 
from those procedures appear summarized in Fig.~\ref{fig:alpha_s}.(a).

\section{Lattice computation of the coupling in the Taylor scheme}

In ref.~\cite{Blossier:2010we}, we calculate the strong coupling constant from the ghost-gluon vertex 
through 
\beq\label{alpha}
\alpha_T(\mu^2) \equiv \frac{g^2_T(\mu^2)}{4 \pi}=  \ \lim_{\Lambda \to \infty}
\frac{g_0^2(\Lambda^2)}{4 \pi} G(\mu^2,\Lambda^2) F^{2}(\mu^2,\Lambda^2) \ ,
\eeq
where $F$ and $G$ are the ghost and gluon dressing functions and 
$\Lambda=a^{-1}(\beta)$ is the regularisation cut-off. This coupling 
is renormalized in the MOM Taylor scheme, where the ghost-gluon vertex 
is finite and the only form factor surviving goes to 1~\cite{Taylor} 
because the incoming ghost momentum is taken to vanish. Here $g_0$ is the bare strong
coupling and $\mu$ the renormalization scale.
This definition can be used in a lattice determination and is to be compared with a theoretical 
formula in order to extract $\Lambda_{\rm QCD}$:
\beq\label{alphahNPt}
\alpha_T(\mu^2) =
\alpha^{\rm pert}_T(\mu^2)
\ \left(  1 + \frac{9}{\mu^2} \frac{g^2_T(q_0^2) \langle A^2 \rangle_{R,q_0^2}} {4 (N_C^2-1)}
\right) \ ,
\eeq
where $\alpha^{\rm pert}_T(\mu^2)$ is available at the four-loop 
level~\cite{Blossier:2010we,Chetyrkin00} in PT and, to 
cure the observed mismatch beetween lattice and perturbative determination, a non-perturbative OPE 
correction to the perturbative formula is to be considered. This accounts for the minimal power 
correction associated to the presence of a dimension-two $\VEV{A^2}$ 
condensate~\cite{Boucaud:2000ey,Blossier:2010we}. The  $\Lambda_T$ in the MOM Taylor-scheme 
and the dimension-two gluon condensate are to be obtained from the confrontation, 
over a large momentum window, of \eq{alphahNPt} to 
the lattice data computed from \eq{alpha} and properly cured of lattice artefacts, 
as explained in ~\cite{Boucaud:2000ey}.  Then, we applied this procedure and exploited the ETMC 
lattice configurations~\cite{Baron:2009wt} with $N_f=2$ twisted-mass dynamical quark flavours and, 
after the conversion of $\Lambda_T$ to $\overline{\rm MS}$, obtain (see Fig.~\ref{fig:alpha_s}.b):
\beq\label{final}
\Lams = \left(330 \pm 23\right) \times \frac{0.0801 \ \mbox{\rm fm}}{a(3.9)} \ \mbox{\rm MeV} \ , \, \,
g^2(q_0^2) \langle A^2 \rangle_{q_0} =  \left(2.4 \pm 0.8\right) \times \left(\frac{0.0801 \
\mbox{\rm fm}}{a(3.9)}\right)^2 \ \mbox{\rm GeV}^2 \ ; \
\eeq
where $a(3.9) \ = \ 0.0801(14)~\mbox{\rm fm}$~\cite{Baron:2009wt}. Of course, with only two sea quark flavours, 
the computation of $\alpha_{\overline{\rm MS}}(m_Z)$ is still inaccesible. A computation from $N_f=4$ lattice 
simulations is now in progress.

\section{Conclusions}

We shortly reported on some very recent computation of $\alpha_{\overline{\rm MS}}(m_Z)$ from $N_f=2+1$ 
lattice simulations, allowing a matching from $N_f=3$ to $N_f=4$ wich uses PT and neglects the impact of 
the still important non-perturbative impact at the charm quark mass. Then, we also reported on the computation 
of the Strong coupling from the ghost-gluon vertex over a large momentum window, which reveals the impact 
of the non-perturbative effects at energies of the order $O(1)$ GeV and leads to an estimate of $\Lams$ 
for $N_f=2$ consistent with other independent computations.

\end{document}